\documentclass[12pt]{article}          
\usepackage{times}                               
\usepackage {subfigure}
\usepackage {multirow}
\usepackage{latexsym}
\usepackage{amsmath}
\usepackage{amssymb}
\usepackage{graphicx}
\usepackage{setspace}
\usepackage{amsmath}
\usepackage{ctable}
\usepackage{epsfig}
\usepackage{textcomp}
\usepackage{multicol}
\usepackage{mdwlist}
\usepackage{url}
\usepackage{abstract}

\usepackage[hmargin=1.3cm,vmargin=3.5cm]{geometry}

\newcommand{\REM}[1]{}  
\newcounter{step}

\long\def\comment#1{}

\usepackage{sectsty}
\sectionfont{\large}
\begin{document}

\onecolumn   
\title{Instant Replay: Investigating statistical analysis in sports.}
\author{
Gagan Sidhu, \\
Department of Computing Science, \\
University of Alberta, Edmonton, Canada\\
}

\maketitle


%

Technology has had an unquestionable impact on the way people watch sports. Along with this technological evolution has come a higher standard to ensure a good viewing experience for the casual sports fan. It can be argued that the pervasion of statistical analysis in sports serves to satiate the fan's desire for detailed sports statistics. The goal of statistical analysis in sports is a simple one: to eliminate subjective analysis. In this paper, we review previous work that attempts to analyze various aspects in sports by using ideas from Markov Chains, Bayesian Inference and Markov Chain Monte Carlo (MCMC) methods. The unifying goal of these works is to achieve an accurate representation of the player's ability, the sport, or the environmental effects on the player's performance. With the prevalence of cheap computation, it is possible that using techniques in Artificial Intelligence could improve the result of statistical analysis in sport. This is best illustrated when evaluating football using Neuro Dynamic Programming, a Control Theory paradigm heavily based on theory in Stochastic processes. The results from this method suggest that statistical analysis in sports may benefit from using ideas from the area of Control Theory or Machine Learning.

\noindent\textsc{Keywords}: {dynamic programming, markov, baseball, basketball, football}

\twocolumn

\section{Introduction}\label{intro}\
\par	Over time, statistics and sports have become synonymous with each other. This synonymity between sport and statistics can be best exemplified when observing the winner and loser of a game. A team only wins by outscoring its opponent. However, in order to outscore an opponent, additional contributions from non-scoring players on the team are required. As a result, to reflect each player's contribution to the team statistical categories have been added. Over time, these player contributions have been further refined to best reflect $\mathit{overall}$ performance. 
\par An example of this is in basketball when a player retrieves the ball after a missed shot. Such a retrieval is called a rebound. This statistic has been further refined in to two categories: defensive rebound and offensive rebound. A defensive rebound is when the opposing team's player shoots and misses and the defending team's player successfully retrieves the ball. An offensive rebound is when a player on the team misses the shot, but a player on the same team manages to retrieve the ball. This refinement allows observers to assess a player's rebounding abilities when playing defense or offense.
	
\par	This refinement provided by statistical analysis seeks to accurately and impartially evaluate player and/or team talent. What can make this analysis difficult is the team-oriented nature of sports, because player performance can be heavily impacted by the team's performance. Various authors have attempted to use different techniques to assess a player's ability, independent from their team, with mixed results.
\par	 It is also difficult to assess the performance of a team in a hypothesized situation.  Such an example would be investigating the impact of a certain team's batting lineup affecting the outcome (win or loss) of the game. Could there have been a different lineup ordering such that the team could have won the game? As we will see, there has been some work done in attempt to answer this question.
\par The human element, in conjunction with the statistical nature of sports, provides a challenge for any researcher who seeks to investigate consequences of certain phenomena.  The statistical analysis done in sports indirectly tries to assess players' Decision Making when facing uncertainty. Many areas of science are interested in decision making when facing uncertainty, and good results will likely benefit more than just one discipline of science.
\par The goal of this review is to provide researchers unfamiliar with statistical analysis in sports with an understanding of what important research has been conducted in the area over the last four decades. It is assumed the reader is familiar with the methods described in the appendices.
	
\section{Previous Work}\label{prvw} \
	\par There has been a wide array of statistical analysis performed on sports, primarily baseball. In this section, we discuss the important results in the past four decades. These results provide an idea of how statistical analysis can be used to evaluate desired properties of either players, teams or the game itself.
\
\subsection{Offensive Earned Run Average for Baseball}
\	 In the introduction, it was mentioned that a player's performance can be significantly impacted by their team. In baseball, this is especially true with statistics such as Runs Batted In (RBI), which is credited to a batter when the outcome of their at-bat scores a run. This means that if the outcome was a fly out and a run is scored, the batter is credited with a RBI even if the base runner had the speed to out-run the throw to home plate. This shows that there exist player statistics in baseball that depend on the team's overall abilities. 
\par Offensive Earned Run Average (OERA) was created by Cover and Keilers (1977) to accurately assess a player's offensive output to avoid situations such as the above. That is, they wanted to create a metric that was $\mathit{independent}$ from the team's performance. The central idea behind OERA is $\mathit{personal\; innings}$, which are defined as innings where the player bats at every position in the lineup. The provided example was one where a batter who starts his career with the at-bat sequence ``single, out, double, out , walk, walk, homerun, out" generating five runs in this $\mathit{personal\;inning}$. The claim was that OERA is a measure of ``batter effectiveness", with units of measurement being the expected number of runs scored per game.
\par The idea of personal innings is effective when the goal is to measure a player's individual offensive contribution to the team. A player who possesses a high batting average, on base percentage and slugging percentage will generate more runs than a player with the same on base percentage, slugging percentage but a lower batting average. 
By measuring batting effectiveness in such a manner, a player's OERA depends on their ability to reach base and advance/score baserunners. 

%

The batter's cumulative statistics are used to compute the probability of achieving any of the six hitting outcomes. The six outcomes are: strike out, walk, single, double,triple and home run, and their probabilities are denoted as $p_0,p_B, p_1,p_2,p_3,p_4$, respectively. The expected number of runs is generated by using these probabilities, averaged over all the possible sequences of hitting performances. 

\par The model to represent the game state consists of $8 \times 3 = 24$ states. This is because when there are zero, one, or two, there are exactly $2^3$ different positions that the base-runners can take. 
Using a three digit binary sequence, where the digit position denotes a base runner at that base, the eight states for each out become clear (000,001,010,100,101,011,110,111). The third out state is ignored since it signifies the end of the inning.

\par Let $\mathbb{E}(s)$ represent the expected number of runs scored in an inning when starting in state $s$. Since there is a Markovian Recurrence for $\mathbb{E}(s)$, it can be written as:
\begin{equation} \begin{split}
\mathbb{E}(s)  = \sum_{H} p_H (\mathbb{E}(f(H,s)) + R(H,s)) =\\ \sum_{s'} p(s'|s)\mathbb{E}(s') + R(s) 
\end{split} \end{equation}

\noindent where $H \in \{0,B,1,2,3,4\}$ represent the hitting outcomes, and $s \in \{0,\ldots,24\}$ are the possible states.   The resulting state $s'$ from a hit is determined by function $s' = f(H,s)$. The number of runs scored by the hit is denoted by $R(H,s)$. A state-transition function is defined as $p(s' | s) = \sum_{H:f(H,s)=s'} p_H$.

This expectation also has an equivalent representation using the theory of Absorbing Markov Chains:
\begin{equation} E = (I-Q)^{-1} R \end{equation}
Where Q is the $24 \times 24$ representing the 24 non absorbing states previously mentioned, R and E are $24 \times 1$ vectors representing the expected number of runs and runs in those 24 states. As an example, the first entry in E, E(1), represents the expected number of runs earned with no outs and no men on base.
Using the probabilities $(p_0,p_B, p_1,p_2,p_3,p_4)$, the negative binomial distribution can be used to calculate the probabilities required to find the expected number of batters in an inning. Specifically:
\begin{equation} \mathbb{P}(N=i) = {i-1\choose 2} p_0^3q_0^{i-3}\end{equation}
where $q_0$ is defined as the on base percentage (OBP), expressed as $q_0 = 1 - p_0$.~\footnote{Note that this is the case because every other hitting outcome consists of the batter getting on base, and so subtracting $p_0$ from 1 gives us the probability the player will get on base.} Therefore, taking the expectation of the above expression gives $\mathbb{E}(N) = {3}/{p_0}$, the expected number of batters in an inning. Note that the OBP, $q_0$, for a given type of hitter will differ. To illustrate the meaning of ``type of hitter", assume we classify batters as either singles, doubles, triples or home run hitters. Then the OBP for a home run hitter would be $q_0 = 1 - p_0$, where $p_1=p_2=p_3 = 0$ since this hitter $\mathit{only}$ hits home runs. 
\par The number of runs generated by a home run hitter would be $R_4 = ({3}/{p_0}) - 3 = {3q_0}/{p_0}$, which is the expected number of batters in an inning minus the number of outs (three). The idea is similar for the other types of hitters. Note that $R_i$ denotes the expected number of runs earned for a hit of type $i$, where $i \in {B,1,2,3,4}$.

\par Using the above information, the general case of ``pure-hitters" can be defined. Assume that $N$ is the random variable for the number of batters in an inning. For each type of hitter, there exists a minimum number of runners required prior to scoring a run; as a result Covers and Keilers (1977) define, for any real number $t$, $(N-t)^+ = N-t$ if $N- t \geq 0$ and $(N-t)^+ = 0$ otherwise. Since will be at $\mathit{least}$ three players in each inning, $t \geq 3$ because at least three batters are required to end an inning, assuming each of their at-bat outcomes resulted in an out.
\par This means a home run hitter's expected number of runs is $R_4 = E(N-3)^+$ because they have three outs, but do not require any men on base to score. Similarly for a singles hitter $R_1 = E(N-5)^+$ because there must be two men on base (one on first base and one on second base due to the assumptions) in addition to the three outs before he can score. The doubles and triples hitter have the same number of players on base required, one, which gives us $R_2 = R_3 = E(N-4)^+$. For the all walks hitter, they need three men on base in addition to three outs and therefore $R_B = E(N-6)^+$. Using this information, Covers and Keilers (1977) give the final expressions: 
\begin{equation} \begin{split} R_B  = 3/p_0 - 6 + 3p_0^3(1+2q_0+2q_0^2) \\ R_1 = 3/p_0 - 5 + 3p_0^3q_0+2p_0^3\;\;\;\;\;\;\;\;\;\;\;\; \\ R_2=R_3 = 3/p_0 - 4 + p_0^3 \;\;\;\;\;\;\;\;\;\;\;\;\;\;\;\;\;\; \\ R_4 = 3/p_0 - 3 \;\;\;\;\;\;\;\;\;\;\;\;\;\;\;\;\;\;\;\;\;\;\;\;\;\;\;\;\;\;\;\;\;\;\; \end{split} \end{equation}

Where the player's OERA is the sum of these expressions. 

\par The OERA calculation is airtight in the sense that it uses the most independent player statistic to calculate this production, which is the batting average. The drawback is $\mathit{how}$ OERA uses this statistic. It does not factor in the league-wide average in order to get a true measure of a player's batting effectiveness. We give the following scenario to illustrate this problem.

\par Consider a batter who's career batting average was 0.270 with an OBP of .330, where the league-wide batting average was 0.285 with an OBP of 0.350 during his career. Now consider a player from the ``deadball era" with a career batting average of .230 and OBP of 0.3, where the league wide batting average was 0.21 with an OBP of 0.25. The second player was clearly better with respect to the league performance, but his OERA will be much lower than the first player. However, is it fair to say that the second player wasn't an effective batter? It can be argued that the second player's batting effectiveness was superior to the first player's given that the second player was above league average and the first player was not.

\par The above scenario outlines a rather large caveat in the OERA calculations. Historically, baseball has shown shown fluctuations with respect to average number of runs scored in a game, and this affects OERA since it doesn't give us a $\mathit{relative}$ measure of players with respect to their peers.

\subsection{Composite Batter Index (CBI)}
\par	Like OERA and the Scoring Index, CBI attempts to quantify a player's offensive abilities. A distinguishing and interesting property of CBI when compared to OERA is that CBI is  a $\mathit{relative}$ measure, meaning it attempts to gauge a player's production with respect to the entire league. It was developed by Anderson et al (1997) using Data Envelopment Analysis. Since CBI is a $\mathit{relative}$ measure, it allows the metric to be insensitive to league-wide changes such as poorer pitching, rule changes, park changes, or increases/decreases in league averages. 
\par 	The CBI model has one input with five outputs. This single input is plate appearances, which contains the official number of at-bats plus the number of walks. Sacrifice flies/bunts and being hit by a pitch are ignored in the calculations. The output, Y, consists of the number of walks, singles, doubles, triples and home runs. 
\par Because CBI relies on a technique called Data Envelopment Analysis, a standard linear programming formulation is used:
\begin{equation}
\begin{split}
\mathrm{minimize}\;\;\; \Theta,\;\;\;\;\;\;\;\;\;\;\;\;\;\;\;\;\;\;\;\;\; \\
\mathrm{subject\; to}\;\; Y\lambda \geq Y_0,\;\;\;\;\;\;\;\;\;\; \\
\Theta X_0 \geq X'\lambda \;\;\;\;\;\;\; \\
\Theta \mathrm{free}, \;\;\; \lambda \geq 0  \;\;\;\;\\
\end{split}
\end{equation}

where $\Theta$ is a score that measures the productivity of the player relative to the rest of the league, and is in the range of 0 and 1.0. This means if a player has a $\Theta = 0.8$, then some hitter (or combination of hitters) could have produced at least the same amount of each type of hit in 20\% fewer plate appearances. A value of $\Theta = 1.0$ implies that the player is a league leader because they can't be surpassed by any combination of players in equal or less plate appearances. $\lambda$ is a vector of virtual multipliers that describe the combination of league leaders that are equal or greater than the player studied.
\par Anderson et al's (1997) initial results showed that players were able to achieve league-leader status only on the basis of being able to obtain singles or walks. After further examination they concluded that these evaluations were unreasonable because there were certain players who hit enough longer hits to compensate for the deficit in shorter hits, and consequently should surpass these short hitting league leaders. 
\par They argue that a player who has a larger number of longer hits can elect to stop on first base during these hits, instead of continuing to second or third base. By this argument, Anderson et al (1997) conclude that there is a dominance relationship among the types of hit. This dominance would allow single base hits to be the sum of singles, doubles, triples, and home runs, double base hits to be the sum of doubles, triples, and home runs, triple base hits to be the sum of triples and home runs, and walks to be the sum of walks, singles, doubles, triples and home runs. This ``dominance transformation" was performed as a pre-processing step. All analyses for the remaining part of their paper involved using this transformation.
\par The outcome of this analysis indicated there were some (statistically unproven) trends in the historical data that was used.  Anderson et al (1997) state that there is a trend towards a higher league-wide CBI which implies that batting skill is becoming more uniformly distributed. With the recent divulging of the ``steroid era" in baseball, this seems to be a reasonable assumption as players are continually trying to improve their performance. Also mentioned was the increase of league leaders in each year when looking at the raw numbers: there were six occurrences when CBI league leaders included ten players or more, but only one of these was before 1975. This lead Anderson et al (1997) to conclude that it is more difficult to dominate a league than it was in the past. Lastly, they mention that the proportion of players with high CBI scores has increased and the number of players with low CBI scores has decreased.
\par Treating a player's offensive performance as a linear program is an intriguing concept, however the methodology itself must be questioned. While the idea of a relative measure is most definitely an enticing one, is an optimization technique the best way to derive such a measure? Constraint programming's versatility cannot be argued, but the application to a sport in order to measure a player's offensive performance relative to the league is difficult.

\par 	As we have seen already, with the use of statistics, it is possible create effective player metrics that measure a certain aspect of baseball. CBI and OERA seek to measure offensive performance in different ways, consequently leading to very different models. Each of these models sought to evaluate $\mathit{individual}$ performance, but what about team performance? OERA's model consisted of 24 states, where the expected number of runs from each state could be calculated with a Markovian Recurrence. It seems logical that with a few modifications that OERA's model can be used for $\mathit{each}$ player on a team, allowing team performance to be measured.

\subsection{Markov Chain Approach to Baseball} 
\subsubsection{Method} \
\par The Markov chain method proposed by Bukiet et al. (1997) sought to evaluate the the baseball team performance and the effect of a player on team performance. They noticed that ``run-production" models like OERA did not lend insight to $\mathit{team}$ performance.
\par Of specific interest to them was the the influence of batting order on a team's performance. They also use this method to approximate the expected number of wins in a season, the number of runs of scored in an inning, and the influence of trading a player on the team wins.  If a team wins a game, they must outscore their opponent. The best batting order is the one that produces the most runs. It is easy to see then that this model relies heavily on run production to evaluate these different aspects. This can be a potential issue, since there exist teams (such as the 2010 San Francisco Giants) which rely on defense and pitching, instead of hitting, to win games.

\par It was mentioned that with slight modifications, the OERA model could be used to measure team performance. Since team performance is being measured, the three-out state serves as an absorbing state denoting the end of a team's inning. Consequently, each player posseses a $25 \times 25$ transition matrix. Each entry in this matrix contains the probability for this player, in a single at bat, to change the game state to any other state. Because of low data availability,  simple statistical models were used with the available data for each type of hit to fill this matrix. 
\par Generating data for probabilities of each hitting outcome for a player can be hazardous. In contrast, OERA did not have this issue since only six hitting outcomes were required, and the baserunners' position did not affect a player's hitting probabilities. In this case, the six hitting probabilities for each player must be generated for different baserunner positions. Generating data in such a manner may cause a player's $25\times25$ transition matrix to be unrepresentative of their real-world performance.
\par An advantage of this model is that the situational data can be implemented into the transition matrix with ease. Consider a batter who is twice as likely to hit a home run with the bases loaded. By multiplying all entries in the transition matrix that correspond to a transition to a state with no runners on base and the same number of outs, the probability of a player hitting a home run in bases loaded has now been doubled.
\par The $25 \times 25$ (block) transition matrix for every player has the following form:
\begin{equation} \mathbf{P} = \left( \begin{array}{cccc}
A_0 & B_0 & C_0 & D_0 \\
0 & A_1 & B_1 & E_1 \\
0 & 0 & A_2 & F_2 \\
0 & 0 & 0 & 1	\end{array} \right)
\end{equation}

Where all A, B, and C matrices are $8 \times 8$ block matrices, and $D_0, E_1, F_2$ are $8\times 1$column vectors.  The $8 \times 8$ block matrices represent the eight possible states of base runners; more specifically A represents the events which do not increase the number of outs in the inning, B represents the events for which the outs increase by one but do not end in three outs, and the C matrix represents the events that result in a double play (from no outs to two outs). The rows of each of these matrices represent the transition probabilities $\mathit{from}$ these states. Therefore the $i$th row and $j$th row of the matrices is the transition from the $i$th state to the $j$th state.
\par From this information, we know that there are certain transitions where one can infer that a run is scored.  Consider the transition from the state with zero outs and base runners on first and second base to a state with zero outs and a base runner on second base; this can only occur when two runners score. The runs scored from these transitions are used to calculate the run distribution for each team.
\par To calculate the distribution of runs in the game, the probability of scoring any number of runs until the current at bat is calculated.  For a single inning, this calculation involves a $1 \times 25$ vector $\mathbf{u}_0$ whose first entry is one, and the remaining 24 entries are zero. This vector represents with state with no outs, and no one on base. Similarly, $\mathbf{u}_n$ represents the situation where $n$ batters have already had their turn batting. Since $n$ batters have already gone, then it is the turn of the $n+1$ batter. Therefore  $\mathbf{U}_n \times \mathbf{P}_{n+1}$, where $\mathbf{P}_{n+1}$ is the transition matrix of the batter whose turn it is, a probability distribution of the states in the inning after $n+1$ batters. In order to keep track of the number of runs scored until this point in the inning, a $21 \times 25$ matrix $\mathbf{U}_0$ which has 21 rows to represent zero to 20 runs (the first row is zero runs, the last row is 20 runs), and 25 columns representing the current state of the inning, is maintained. 

\par When there is a transition that causes a run to score, the probability of this outcome is propagated to $\mathbf{U}_{n+1}$. This relatively simple representation allows the distribution of runs to be calculated after any number of batters as follows:
\begin{equation}
\mathbf{U}_0 \mathbf{P}_1 \mathbf{P}_2 \mathbf{P}_3 \ldots \mathbf{P}_9 \mathbf{P}_1 \mathbf{P}_2 \ldots
\end{equation}
where each subscript of $\mathbf{P}$ represents one of nine players. 
Because this is for a single inning, $\mathbf{U}_n$ is given nine times as many rows, where each row set of 21 rows represents the number of runs in an inning. When an inning ends, the results are propagated to the next 21 rows and 24 columns (the twenty fifth column is absorbing, and no scoring is done there). This computation is continued until the  probability that 27 outs have occurred is greater than 0.999.
\par It is apparent that this model relies heavily on generated data and an iterative stopping condition. When these two ideas are combined, the result may be completely unrepresentative of the players and teams being evaluated. Using sample data to generate probabilities of hitting outcomes in specific situations alone can cause results to be unrepresentative; using this data in an iterative computation that determines the stopping criteria will yield results that are unrepresentative of the team and player.

\subsubsection{Scoring Index}\

\par One of the primary goals in developing the Markov Chain approach was to compute the near-optimal batting order for a baseball team. With the Markov framework established as above, they also needed a way to rank a player's offensive ability. Instead of using OERA, they used a similar metric proposed by D'Esopo and Lefkowitz called the Scoring Index because of the similarity in run production calculation between the Markov Approach and Scoring Index. 
\par The Scoring Index is similar to OERA in sense that it uses nine copies of the batter in the lineup to calculate the offensive production, however there is ranking among these nine copies using a deterministic model of runner advancement. Also similar to OERA, the Scoring Index contains similar assumptions to ensure the computation is consistent, deterministic and representative of the player's offensive capabilities. The distinguishing factor is that it uses one inning of the Markov approach outlined above. 
And the transition matrix $\mathbf{P}$ for the scoring index is also $25 \times 25$, defined as:
\begin{equation} \mathbf{P} = \left( \begin{array}{cccc}
A & B & 0 & 0 \\
0 & A & B & 0 \\
0 & 0 & A & F \\
0 & 0 & 0 & 1	\end{array} \right)
\end{equation}
Inspecting this $P$, it is apparent that it is different from the $P$ given in the Markov Chain section. This is due to block matrices $C_0, D_0$ and $E_0$ becoming zero because the Scoring Index ignores double and triple plays. The off-diagonal entries of block matrix B are zero because runners cannot advance on an out. Just as OERA has some assumptions that may affect the calculation, the Scoring Index does as well. The claim was that these inaccuracies should ``somewhat offset each other".

A and B are block matrices with the following structure:
\begin{equation} \scriptsize A=\left( \begin{array}{cccccccc}
P_H & {P_S + P_W} & P_D & P_T & 0 & 0 & 0 & 0 \\
P_H & 0 & 0 & P_T & {P_S + P_W} & 0 & P_D & 0 \\
P_H & {P_S} & P_D & P_T & P_W & 0 & 0 & 0\\
P_H & {P_S} & P_D & P_T & 0 & P_W & 0 & 0\\
P_H & 0 & 0 & P_T & {P_S} & 0 & P_D & P_W \\
P_H & 0 & 0 & P_T & {P_S} & 0 & P_D & P_W \\
P_H & {P_S} & P_D & P_T & 0 & 0 & 0 & P_W\\
P_H & 0 & 0 & P_T & {P_S} & 0 & P_D & P_W \\
	\end{array} \right)
\end{equation}

\begin{equation}
B = P_{out}I
\end{equation}
where the probabilities of getting a walk, single, double, triple, home run or out are denoted as $P_W, P_S, P_D, P_T, P_H$, and $P_{out}$, respectively. I is an $8 \times 8$ identity matrix. F is a $8 \times 1$ vector with the entries: $F = (P_{out}, \ldots, P_{out})^T$. 

\par We can use the theory of Absorbing Markov Chains to calculate the time to the three out state, which is the absorption state. The $(i,j)th$ entry of matrix $(I-Q)^{-1}$ gives the expected number of visits from state i to state j prior to being absorbed into the three out state. This is referred to as the ``expected absorption time". The expected absorption time from state $i$ can be calculated by : 
\begin{equation}
\mathbb{E}_i (T)  = \sum_{j=1}^{24} (I-Q)_{ij}^{-1}
\end{equation}
and for the Scoring Index model:

\begin{equation} Q = \left( \begin{array}{ccc}
A & B & 0 \\
0 & A & B \\
0 & 0 & A \end{array} \right)
\end{equation}
which results in $(I-Q)^{-1}$ having the following structure:
\begin{equation} (I-Q)^{-1} = \left( \begin{array}{ccc}
R & RBR & RBRBR \\
0 & R & RBR \\
0 & 0 & R \end{array} \right)
\end{equation}
where R = $(I-A)^{-1}$.
\subsubsection{Near-Optimal Batting Order}\
\par Using the Scoring Index as a way to rank each player's offensive ability, the optimal batting order was calculated with three different algorithms. These three algorithms focused on computational efficiency, since CPU time was not cheap at the time the article was written. As a result, the differences in these algorithms is omitted since most modern day computers can calculate the 9! (362,880) permutations relatively quickly. It was observed that the lineup that produces the most runs was also the one with the most wins, which should follow naturally if the batting order is indeed optimal.

%
By comparing the worst lineup to the team's near-optimal one, the results suggested the following criteria when constructing a team's batting lineup:
\begin{enumerate*}
\item The batter with the highest scoring index should bat second, third or fourth.
\item The batter with the second highest scoring index should bat between the first and fifth positions.
\item The batters with the third and fourth best scoring indices should bat between the first and sixth positions.
\item The batter with the fifth highest scoring index should bat first, second or between the fifth and seventh positions.
\item The sixth best batter can bat in any position except eighth or ninth.
\item The batter with the seventh highest scoring index can bat either first, between sixth through ninth positions.
\item The batters with the lowest and second-lowest scoring indices should bat in the last three positions.
\item Either the batter with the second or third highest scoring index should bat right after the best batter.
\item The batter with the lowest scoring index should be four to six positions after the best batter.
\item The batter with the second lowest scoring index should be four to seven batters aft¶er the best batter.
\end{enumerate*}

\par These ten possible criteria seem very reasonable, however there are no conclusive results showing that lineups following this criteria are optimal in comparison to the team's regular lineup, or whether this near-optimal lineup yields more runs (and consequently more wins) than the regular lineup. Further evaluation of this method is required; specifically there should be a comparison between the near optimal lineup of the team versus the regular lineup to determine whether this lineup will yield more wins.  There should also be testing on more recent data, since the analysis was done on a single year of data (from 1989). 
\subsubsection{Conclusion}
\par This article was the first to use an offensive metric (Scoring Index, similar to OERA) for nine players on the same team to evaluate the team performance. As mentioned earlier, the methodology in this article may have affected the significance of the results. With the recent abundance of data provided by MLB's GameDay system, it should be possible to calculate each player's hitting probabilities in specific situations. Using such data may yield a much more promising result if this experiment is repeated. 
\par It is apparent that evaluating team performance is difficult if there isn't an accurate representation of each player. So far we have only considered the offensive output of a player while assuming their performance is not affected by the environment. However, statistics show that players performance is affected by specific environments. When considering a player's performance, whether it is offensive or defensive, their performance in different conditions must be considered if the team wants to maximize the probability of winning each game. 

\subsection{Breakdown Statistics}
	Albert (1994) broke down players' annual statistics in order to test whether batter performance can be affected by certain situations. As an example: if a batter bats poorer when he's not playing on home field, is it because of the playing surface or the fact that he isn't comfortable playing away from home field? In order to do this, eight common situations in which players' averages could be affected were evaluated:

\begin{enumerate*}
\item Opposite side versus the same side. This means that when a pitcher is right-handed and throwing to a right-handed batter, this is the ``same side". When a pitcher is left-handed throwing to a right-handed batter, it is the ``opposite side".
\item Whether a pitcher is a ``groundball pitcher" or ``flyball pitcher". This means that this pitcher relies on batters hitting ground balls (or fly balls) to the pitcher's defense in order to achieve an out.
\item Home versus away.
\item The playing surface (grass versus turf)
\item When a batter is ``head in the count" versus two strikes in the count. Being ahead in the count means there are more balls than strikes (exception of three balls and two strikes, which is referred to as a ``full count").
\item Runners in scoring position versus no runners in scoring position and no runners out.
\item Performance in the first ``half" of the season versus the second ``half". The halfway point is determined by the All Star game.
\end{enumerate*}

It has been shown that there is a high correlation between player performance and whether they are playing at home or away from home. In order to evaluate the performance in each of these situations, the number of hits and the at bat for home and away games were recorded for 154 players. For the $i$th player, this can be represented as a $2 \times 2$ contingency table:
\begin{equation}
\begin{tabular}{| c | c | }
  \hline                       
$h_{i1}$ & $o_{i1}$ \\ \hline
$h_{i2}$ & $o_{i2}$ \\
  \hline  
\end{tabular}
\end{equation}
where $h_{i1}, o_{i1}$, and $ab_{i1}$ denote the number of hits, outs, and at bats during home games; $p_{i1}$ and $p_{i2}$  denote the probabilities that the $i$th player gets a hit at home or away, respectively. It is assumed that the batting attempts are independent Bernoulli trials with the associated probabilities of success, and that the number of hits $h_{i1}$ and $h_{i2}$ are independently distributed according to the binomial distributions with the parameters ($ab_{i1}, p_{i1}$) and ($ab_{i2}, p_{i2}$) respectively.
\par This information is transformed to approximate normality by use of the logistic transformation:
\begin{equation} y_{ij} = \log\bigg( \frac{h_{ij}}{o_{ij}} \bigg) \end{equation}
which allows $y_{i1}$ and $y_{i2}$ to be independent normal with each $y_{ij}$ having mean $\mu_{ij}=\log(p_{ij}/(1-p_{ij}))$ and variance $\sigma_{ij}^2 = (ab_{ij}p_{ij}(1-p_{ij}))^{-1}$. If the contingency table is now expressed as $2 \times N$ we can represent every player. With this representation, we know that $y_{ij}$ is the logit of player $i$'s observed batting average in situation $j$. This allows the mean of $y_{ij}$ to be represented as :
\begin{equation} \mu_{ij} = \mathbb{E}(y_{ij}) = \mu_i + \alpha_{ij} \end{equation}
where $\mu_i$ represents player $i$'s hitting ability and $\alpha_{ij}$ represents the situational effect that attempts to measure the change in this player's hitting ability due to situation $j$.

\par The ability parameters $\mu_1,\ldots,\mu_N$ are assigned independently assigned flat noninformative priors as they are nuisance parameters. The reason the ability parameters are considered as nuisance is because the goal is to measure the $\mathit{effect}$ of the environment on the player's hitting ability, which is given by the situational effects . Since the situational effects $\alpha_1,\ldots,\alpha_N$ are of interest, it is assumed a priori that $\alpha_1,\ldots,\alpha_N$ are independently distributed from some common population $\pi(\alpha)$. 
\par The prior distribution used was a $t$ distribution with mean $\mu_{\alpha}$, scale parameter $\sigma_{\alpha}$ and a $\nu=4$ degrees of freedom. In order to reflect the lack of knowledge about the size of the situational effect, $\mu_{\alpha}$ is assigned a noninformative prior. For $\sigma_{\alpha}^2$, an informative prior is constructed using the home/away variable as the the representative for the situational variables. The prior of $\sigma_{\alpha}^2$ is based upon a posterior analysis of the home/away situational variable in previous seasons. This prior distribution is used in the posterior analysis for every situational variable.

\par The results used a Gibbs sampler to simulate the posterior distributions. For a sample size of 1000, a posterior distribution was obtained with parameters $(\{\mu_i\}, \{\alpha_i\}, \mu_\alpha, \sigma_\alpha^2 \}$. These parameters were used to estimate the functions of the parameters of interest, primarily the $\alpha_1,\ldots,\alpha_N$. The final data showed that the spread of each population for a given situation was roughly the same, allowing the differences between situations to be explained in terms of a shift. This is  exemplified by the ``groundball-flyball" effects, which are approximately ten batting average points higher than the ``day-night" effects.

\par While nearly all of the situational effects being measured did not reveal a significant difference in batting average, the effect of pitch count did. Specifically, when a batter was ``behind" in the pitch count, they hit 123 average points lower than when they were ``ahead" in the count. A less significant result was that batters also hit 20 average points higher when facing a pitcher that was on the opposite side, 11 points higher when they were facing a groundball pitcher and eight points higher when at home.  However, a caveat to these results is that these situational patterns are only apparent for a $\mathit{group}$ of players, and not individuals. This was evidenced by nine players who showed extreme estimated effects for one season, but many exhibited the opposite sign for the previous four years. This inconsistency led to the suggestion that season performance might be an imperfect form of measurement of situational abilities, and that play by play data might be more helpful. 
\par There is yet to exist any published work which statistically analyzes baseball at the pitch by pitch level. It was mentioned earlier that the depth,availability and accessibility of baseball data recently suggests that this topic should be re-visited.

\par What is most enticing about this article is the level at which it attempts to make inference. It is likely that if data was obtained at the pitch by pitch level, the situational effects could be measured at the individual level. The importance of evaluating situational effects in sports cannot be understated, since performance can be significantly impacted depending upon the situation and player. A question that arises from these results is whether it is possible to simply take account of these situational effects, and then attempt to evaluate a player's performance, rather than trying to measure the magnitude of this effect.  

\par	The OERA/Markov Chain approach both rely on exploiting the Markov Property, whereas Albert (1994) 's evaluation of breakdown statistics relies on Bayesian Inference. Is it possible that a combination of these two techniques can result in a better evaluation of a desired phenomena? The next two articles rely heavily on Markov Chain Monte Carlo (MCMC) methods to generate the sample points that are used in their analysis. Particularly interesting is how each article analyzes a different sport, but both use similar techniques to obtain their results.
	
\subsection{Spatial Analysis of Shot Chart Data in Basketball}\
\par	Statistical analysis in basketball is difficult, largely due to the limited amount of information that is given in the data. If we compare the analysis of basketball to baseball, what makes baseball ``easier" to analyze statistically than basketball is the low dependence on spatial information of player movement. In basketball, the way a player moves up and down the court impacts his shot selection and his decision making, because they are allowed to take shots from anywhere. Contrast this to baseball, where, upon successfully hitting the ball for a base-hit, the player must run on a fixed (base) path to reach base. 
\par Reich et al. (2006) used basketball shot chart data in attempt to infer a player abilities and tendencies so that this player can be played optimally when defending. Namely, they attempted to model the player's shot frequency and location. The player used for analysis was Sam Cassell of the Minnesota Timberwolves during his 2003-2004 season in the NBA. The data was about 90 $\%$ complete, since six games were missing and a few shots were not recorded on the shot chart. 
\par The analysis used the shot location, angle from the center of the basket, time between last shot attempt (game clock time), shot outcome (make or miss) to compute ten covariates that affected the player's shot selection. The purpose of these covariates were to account for different situations when the player was on the floor. That is, certain covariates accounted for the change in a player's performance. It should become apparent to the reader that the function of these covariates is equivalent to the situational effects in Albert's (1994) article.

\par Lastly, shot selection was defined as the location, success and frequency of shots taken, each of which were calculated separately. The shot location was converted from euclidean coordinates to polar coordinates so the relationship between the angle and distance at which the player took a shot could be explored.
\subsubsection{Shot Frequency}\
\par To model the player's shot frequency, Reich et al (2006) calculated the median time between each shot attempt when the player was in the game and explored how this value was affected by the ten selected covariates denoted $X_k,\; k \in \{1,\ldots,10\}$. One limiting factor that should be mentioned is that some of the covariates could only be calculated at the time of a shot attempt. The authors attempted to use an error-in-covariates model to correct for this, but the results were not any better.
\par By using multiple linear regression, $$Y = X\theta + b$$ where $Y = \log(T_i)$ was logarithm between the time of shot attempts $i-1$ and $i$, and X was a matrix containing columns for each effect used in calculating the covariates , interaction between select covariates, and the homogenous error variance. By using the inverse gamma distribution $IG(\alpha_1,\alpha_2)$,  with densities proportional to $(\sigma_e^2{-\alpha_1+1})exp(-\alpha_2/\sigma_e^2)$, IG(0.01,0.01) was defined as the prior for error variance,  and diffuse normal priors of mean 0 and variance 1,000 were selected. Their results were analyzed using the posterior means with 95$\%$ confidence intervals for the exponentials of the regression parameters $(\exp{bk})$ (the fitted multiples of the median times between shots relative to the zero condition). 

\subsubsection{Shot Location}\
\par As mentioned earlier, polar coordinates were used for shot location in order to develop a relationship between the distance and angle of a shot location. To analyze shot locations, the court was divided into 11 regions to capture the 11 distinct angles from which the player attempted a shot. This resulted in discretizing the court into an $11\times 11$ grid to make the computation feasible.  An additional cell was added underneath the basket since shot attempts from this location are often after rebounding, or when the player is ``driving to the net" (running directly to the basket in hopes to get a close shot). Because the location of shot was taken as a response to as many as four covariates, and 48 of the 122 cells contained as few as four shot attempts, the neighbouring cells shared information to stabilize the covariate effects. The goal was to find the effect of a certain subset of covariates on each cell where shot attempts were taken.

\par Multinomial regression was used, where $y_i \in \{1,\ldots,p\}$ was the region of the $i$th shot\footnote{$p$, in this case, was 122 for the number of cells in the grid}, and it was assumed that $y_i|\boldsymbol\theta(\boldsymbol\eta_i)$ followed a multinomial distribution $\big(\theta_1(\boldsymbol\eta_i),\ldots,\\ \theta_p(\boldsymbol\eta_i)\big)$ where $\boldsymbol\theta(\boldsymbol\eta_i) = (\theta_1(\boldsymbol\eta_i),\ldots,\theta_p(\boldsymbol\eta_i))$ and $\theta_j(\boldsymbol\eta_i)$ was the probability that the $i$th shot was taken from region $j$. The probability of each cell depends on shot $i$'s $p$-vector of linear predictors $\boldsymbol\eta_i = \log A + \mathbf{x}_i\mathbf{b}$. $\log A$ is a $p$-vector of offset terms  and $\log A_j$ is equal to the area of region $j$ in euclidean distance. This was essential since each of the 11 regions had differing areas. $\mathbf{x}_i$ was a $q$-vector of covariates and because each of the $p$ regions had their own set regression coefficients to measure the $q$ covariates: $\mathbf{b}_{\cdot j} = (b_{1j},\ldots,b_{qj})^T$ for the $j$th region. Because $\mathbf{b}_{\cdot j}$ and $\mathbf{x}_i$'s are related through $\theta_j(\boldsymbol\eta_i)$, a multinomial logit model was used to calculate $\theta_j(\boldsymbol\eta_i)$:
$$\theta_j(\boldsymbol\eta_i) = \frac{\exp(\log A_j + \mathbf{x}_i^T\mathbf{b}_{\cdot j})}{\sum_{l=1}^p \exp(\log A_l + \mathbf{x}_i^T\mathbf{b}_{\cdot l})}$$

\par If $\mathbf{b}_{k\cdot}$ is the spatially varying coefficient of covariant $k$, is assigned a CAR (conditionally autoregressive) prior CAR($\tau_k)$, then $b_{jk}|b_{kj,l\neq j}$ is normally distributed with mean $\bar{b}_{kj}$ and precision (which is inverse variance) $\tau_k m_j$, where $\bar{b}_{jk}$ is the mean of $\mathbf{b}_{k \cdot}$ at region j's $m_j$ neighbors, $\tau_k > 0$ is the degree of smoothing of each $b_{kj}$ towards its respective neighbors.
 
\par In order to share information among cells that was realistic, the idea of $\mathit{distance}$ and $\mathit{angle}$ neighbors were established. An $\mathit{angle}$ neighbour were adjacent cells that were the same distance from the basket , and a $\mathit{distance}$ neighbour were adjacent cells that were the same angle from the basket. The 2NRCAR (two neighbour conditionally autoregressive) model was used to allow different amount of smoothing between the $\mathit{angle}$ and $\mathit{distance}$ by adding parameter $\beta_k \in (0,1)$ which quantifies the amount of ``influence" each type of neighbour has on the respective cells.

Using 2NRCAR($\tau_k,\beta_k$) as a prior, the distribution of $b_{kj}|b_{kl, l \neq j}$ is normally distributed with mean:
\begin{equation}
\begin{split}
\mathbb{E}(b_{kj}|b_{kl, l \neq j}) = \bar{b}_{ajk}\frac{m_{aj}\beta_k}{m_{aj}\beta_k + m_{dj}(1-\beta_k)} +\\
 \bar{b}_{djk}\frac{m_{dj}\beta_k}{m_{aj}\beta_k + m_{dj}(1-\beta_k)}
\end{split}
\end{equation}

\noindent with precision $\tau_k (m_{aj}\beta_k + m_{dj}(1-\beta_k))$, and similar to the CAR model, $\bar{b}_{akj}$ and $\bar{b}_{dkj}$ are the mean of $\mathbf{b}_{k\cdot}$ for region $j$'s $m_{dj}$ and $m_{aj}$ distance and angle neighbors, respectively. With the above formulation,$b_{kj}|b_{kj,l\neq j}$ to have a weighted average of $\bar{b}_{akj}$ and $\bar{b}_{dkj}$, and $\beta_k$ determines the weight of each neighbor type with $\tau_k$ controlling the smoothness of $\mathbf{b}_{k\cdot}$. It should be intuitive that $\beta_k > 0.5$ smoothes angle neighbors more than distance neighbors, and $\beta_k < 0.5$ smooths distances neighbors more than angle neighbors.
\par Sharing information by use of smoothing ensures that the computation remains numerically stable. The elegance of the above formulation lies in the fact that Reich et al. (2006) distinguished neighbouring cells in two different manners. By doing so, smoothing could be performed with on either distance or angle neighbours if the respective neighbours were suffering from data sparsity. Such a technique is beneficial in alleviating the issue of data sparsity given a limited amount of data.

\subsubsection{Shot Success}\
\par Assuming that the shot location is fixed the probability of a successful shot attempt from each each of the $p$ cells is analyzed. To represent realistic in-game behavior, covariates are allowed to vary over different cells is allowed because the opposition's defensive priorities  may change over time, or the player's team may change their offensive strategy. This can be easily seen when the player's team is up by a significant amount of points, since the intent is to then protect the lead late into a game, which implies less offensive production. The opposition's defensive strategy can change when they are losing the game due to their defense being exposed by a certain player. As a consequence, the defense would then put priority guarding this player, causing this player's covariates to change.
\par To model shooting percentage, separate logistic regressions were each of the $p=122$ locations. The resulting regression parameters for each of these $p$ cells were also smoothed spatially. The probability of making $i$th shot was drawn from a Bernoulli($\pi_i$) distribution, where $z_i \in \{1,0\}$ if the shot was made or missed, and log($\pi_i$) = $\log{\big({\pi_i}/ ({1-\pi_i})\big)} = \mathbf{x_ib_{\cdot y_i }}$ where $y_i$ is the region of the $i$th shot, and $b_{\cdot y_i}$ is the $q$-vector of regression parameters associated with the region $y_i$.

\subsubsection{Conclusion}\
\par In the article, Reich et. al (2006) propose a very interesting methodology using simple spatial data to evaluate a player's shot selection by using covariates that are inferred through the game's data. Both Reich et al. (2006) and Albert (1994) attempt to account for situational effects on the player, but Reich et al (2006) also sought to evaluate their effects on a player's shot selection using sampled data (with the appropriate parameters).

A small drawback is that results given in the article are almost entirely observational, thereby making evaluation of this method difficult. However, this article attempts to address the environmental affects on an athlete, using smoothing techniques which had not been attempted prior. By discretizing the state space into a fine grid, a cell that may have an insufficient amount of data points can be stabilized by use of smoothing on all cells that share some spatial property, such as distance or angle from the basketball net.  It is possible that use of smoothing may improve evaluation of player performance.

\subsection{Bayesball: Evaluating Fielding in Major League Baseball}\

\par	Earlier, we mentioned that statistically analyzing baseball is ``easier" due to the low dependence on spatial information. However there is one case where spatial information is important: Defense. How can we accurately analyze the defensive aspect of baseball? What separates a good defensive player from a poor one? 
	\par Recently, Jensen et al (2009)  developed a defensive metric that sought to evaluate a player's fielding ability with higher accuracy than current metrics such as UZR (Ultimate Zone Rating). UZR discretizes the state space (baseball field) into 54 zones and evaluates the number of successful plays made by a defensive player in each zone. One shortcoming with UZR is the $\mathit{granularity}$ of the state space. It is difficult to accurately measure a player's fielding ability if the area of each zone is large, because the number of plays a fielder makes in a large zone doesn't necessarily lend insight to his mobility and skills. This can be illustrated by comparing two players who play the same defensive position (and therefore start at roughly the same position prior to the catch). If one player catches a ball in play (BIP) little movement (i.e. the BIP is a pop fly), and the other player catches a ball that could be 20-30 feet in front (or behind) him, UZR will evaluate each player's play for that zone as the same.
	\par To alleviate this shortcoming in existing defensive metrics, Jensen et al. (2009) sought to model the success of a fielder on a given BIP as a function of that BIP's location. To do this, a hierarchical bayesian model was fitted to evaluate the individual success of each fielder while sharing information between players at the same position~\cite{bayesball}. The idea of sharing information between players at the same position was motivated by Reich et al (2006)'s work, where the relationship is now based on position of the player. This is a valid relationship because playing the same position suggests a spatial relationship both in distance and angle from the home plate. The values that were obtained by this method were named SAFE, or Spatial Aggregate Fielding Evaluation.
\subsubsection{Models used}\
	\par The models were fitted according to the type of BIP (Flyball, Liner, Grounder) and player position. There are only two models for the BIPs:
\begin{enumerate} 
\item $\mathit{Flyball/liner}$ model:  The $(x,y)$ location of the BIP is set to the location where the ball landed, or where it was caught (if it was caught).  The probability of making a catch was modelled as a function of the distance travelled to catch this ball,  the direction he was travelling, and the velocity of this BIP. The distance travelled incorporates two dimensions because the playing field is a two dimensional plane.
\item $\mathit{Grounder}$ model: The $(x,y)$ location of the BIP is the location where the ball was fielded by either the infielder or outfielder, depending on if the ball made it through the infield. The probability of a fielder making a catch is a function of distance, direction, and the velocity. However, this model's direction is measured by an angle where the BIP was caught and the infielder's starting location. The distance is one dimensional since the infielder now travels the arc length between the location of the catch and the starting position.
\end{enumerate}
	 
	  The reason for this formulation is intuitive: the positions being evaluated have the ability to field flyballs and liners, but infielders almost always field grounders. As a result the outfielders are exempt from the grounder model.\footnote{Pitchers and catchers were ignored due to the lack of data.}The infield positions: 1B (First base), 2B (second base), 3B (third base) and SS (shortstop) had the grounder model fit to their respective position. The remaining three positions: RF (Right Field), CF (Center Field), LF (Left Field), along with the four aforementioned infield positions had the liner/flyball model fit to their positions separately. This resulted in $7\times2  + 4 =18$ models. Because the authors fitted each of the four year's data separately, this resulted in a total of 72 separate models. 
	 \par One large similarity between the models in this article and the previous is their use of ball location. Instead of the location where a shot was taken, it is now where the ball landed (or caught). If a player is responsible for catching a ball and fails to do so, this is treated similarly to a missed shot in the previous article.
	\par Since each of the models are similar, so too are their mathematical formulations:
	
	\begin{enumerate}
	\item $\mathit{Flyball/liner}$ model: For a given fielder $i$, denote the number of BIPs hit when that fielder was playing defense as $n_i$. Since each play's outcome is either a success or failure, then:
	 $$S_{ij}  = \left\{ \begin{array}{ll}
 1 &\mbox{ if the $j$th flyball/liner is caught by player $i$} \\
  0 &\mbox{if the $j$th flyball/liner is not caught by player $i$ }
       \end{array}\right.
       $$
        the observed outcome of a success or failure is modelled as an outcome from a Bernoulli random variable, which has a parameter obtained from an underlying event-specific probability: $$S_{ij} \sim \mathrm{Bernoulli(}p_{ij}).$$
	Each of the Bernoulli probabilities $p_{ij}$ are modelled as a function of distance travelled ($D_{ij}$) to the BIP, an indicator function that represents the direction in which the player was moving ($F_{ij} = 1$ if they're moving forward, $F_{ij} = 0$ if they're moving backwards) and $V_{ij}$ which is the velocity of the BIP:
	\begin{equation}
	\begin{split}
	p_{ij} = \Phi(\beta_{i0} + \beta_{i1}D_{ij} + \beta_{i2}D_{ij}F_{ij} + \\\beta_{i3}D_{ij}V_{ij} + \beta_{i4}D_{ij}V_{ij}F_{ij})\\
	=\Phi(\mathbf{X}_{ij}\cdot \boldsymbol\beta_i)
	\end{split}
	\end{equation} 
	where $\Phi(\cdot)$ is the Gaussian cumulative distribution function, and $\mathbf{X}_{ij}$ is a vector of covariates in the above equation. $\beta_{i0}$ is the parameter that controls the probability of a fielder catching a liner/flyball hit directly at them ($D_{ij}$ = 0), $\beta_{i1}, \beta_{i2}$ are the parameters that control the forward ($\beta_{i1}$) and backward ($\beta_{i2}$) direction of the fielder, and $\beta_{i3}, \beta_{i4}$ are parameters that adjust the probability of a successful catch as a function of velocity. $F_{ij}, D_{ij}$ are both covariates that are functions of the $(x,y)$ coordinates of the BIP. \
	 This model is recognized as the probit regression model with covariate interaction, which permits different probabilities for the same distance travelled leftwards and rightwards.	

\item $\mathit{Grounder}$ model: There are slight changes for this model with respect to the flyball/liner model. Namely, the distance is one dimensional and the direction is now an angle in degrees. The observed outcomes are still Bernoulli realizations as above, however the parameters for the cumulative distribution function change:
\begin{equation}
\begin{split}
p_{ij} = \Phi(\beta_{i0} + \beta_{i1}\theta_{ij} + \beta_{i2}\theta_{ij}L_{ij} + \\\beta_{i3}\theta_{ij}V_{ij} + \beta_{i4}\theta_{ij}V_{ij}L_{ij})\\
	=\Phi(\mathbf{X}_{ij}\cdot \boldsymbol\beta_i)
\end{split}
\end{equation}
where $\theta_{ij}$ represents the angle between the starting position and the location of the BIP, $V_{ij}$ is the velocity of the BIP, and $L_{ij}$ is an indicator function for the direction the fielder has to move towards the BIP ($L_{ij}$ = 1 when moving left, $L_{ij}=0$ moving right).  $\Phi(\cdot)$ is still a Gaussian cumulative distribution function.

\end{enumerate}
\subsubsection{Sharing Information}\

\par With these models being quantified, we now discuss the previously-mentioned idea of sharing information between players.  When the sample size is small, players of the same position have the issue of large variability between their parameter estimates $\beta_i$. If a hierarchical model is used with the assumption that each set of parameter estimates $\beta_i$ are drawn from a common prior distribution, this issue is eliminated.  This is done by drawing each set of player-specific parameters $\beta_i$  from a common distribution shared among players of the same position:
$$\boldsymbol\beta_i \sim \mathrm{Normal}(\boldsymbol\mu, \boldsymbol\Sigma) $$
where $\boldsymbol\mu$ is the $5 \times 1$ vector containing the means and $\boldsymbol\Sigma$ is a $5 \times 5$ prior covariance matrix shared by all players. The components of the $i$th player's parameter estimates $\boldsymbol\beta_i$ are assumed to be independent, even though there is a posterior dependence between the components of this parameter estimate induced by the data.
\par Lastly, a prior distribution must be specified for each position's ``shared player" parameters ($\mu_k,\sigma_k$ : $k=0,\ldots,4$).~\footnote{remembering that only players of the same position share information!}. After extensive investigation, the noninformative distribution was chosen to as: 
$$ p(\mu_k,\sigma_k) \propto 1,\;\;\; k=0,\ldots,4$$ 	
\par If we wish find the unknown parameters {$\boldsymbol\beta$} for $N$ players at a given position, then for each BIP type we will have a $N \times 5$ matrix containing the parameters for each player at the given position, a $5 \times 1$ vector, $\boldsymbol\mu$, which is the vector of parameter means, and $\boldsymbol\sigma^2$, a $5 \times 1$ vector representing the variance among each player's parameters. Thus for each position and BIP-type, the posterior distribution for parameters $\boldsymbol\beta$, $\boldsymbol\mu$, $\boldsymbol\sigma$ are separately estimated,
$$p(\boldsymbol\beta,\boldsymbol\mu,\boldsymbol\sigma^2|\mathbf{S,X}) \propto p(\mathbf{S}|\boldsymbol\beta,\mathbf{X})\cdotp(\boldsymbol\beta|\boldsymbol\mu,\boldsymbol\sigma^2)\cdotp(\boldsymbol\mu,\boldsymbol\sigma^2),$$
where $\mathbf{S}$ contains all of the outcomes  $S_{ij}$ and $\mathbf{X}$ contains all location and velocity covariates $\mathbf{X}_{ij}$. The posterior distribution of the unknown parameters for each position and each type of BIP is estimated using Gibbs sampling.

\par Instead of sharing information across cells, as was the case in the previous article, the information is being shared across $\mathit{players}$. This is because the previous article focused on how the covariates affected shot selection, where shot selection must fall in a certain cell. Just as sharing information with certain neighbouring cells was important in stabilizing the computation in Reich et al's (2006) article, sharing information among players is important since each player has a limited amount of fielding data over a season. By pooling all of this information together for players with the same position, sampling from the appropriate representative distribution, this becomes less of an issue. Both articles rely on using MCMC techniques in conjunction with sharing information to achieve their result.
\subsubsection{Comparing results}\
\par After extensive discussion and evaluation of SAFE, the results~\footnote{Any reader that is curious about the results for which players were evaluated are encouraged to check the article.} for each position are compared to UZR. The difference between SAFE and UZR is that SAFE seeks to evaluates the number of runs that were saved (or cost) by a player's defensive play, while UZR records actual observations. To compare SAFE and UZR, correlation for each player's respective values was attempted, but yielded an inconclusive result. This shouldn't be surprising, because there is no standardized metric that measures a player's defensive play. 
\par To allow further comparison, the assumption that player ability is constant over time was added. This assumption would be supported if the player's value maintained a high consistency across seasons, since the noise in the data would be overcome by the ``true signal". When comparing SAFE's correlation across seasons to UZR's, it did well for the outfield positions but struggled for infield positions such as shortstop and second base.  SAFE and UZR's correlation over these seasons was then averaged over all seven positions, where SAFE had a slightly higher average correlation than that of UZR. This should also not be surprising, since SAFE sought to evaluate defensive play better than UZR by using  an elegant, intuitive and sophisticated representation for evaluation.
\par Another interesting question that can be posed is: Can SAFE's infielder model be adjusted such that it shows a higher level of correlation across seasons? SAFE's results show correlations of (0.525,0.594,0.444,0.503,0.287) for positions (CF,LF,RF,3B,1B) across seasons, but a meagre (0.051,-0.03) for (2B, and SS) respectively. The intrigue behind this lies within the fact that the SS and 2B positions are closer than any other two defensive positions on the field. The distance and angle for which a 2B might need to move to catch a BIP to his right might depend on the calibre of the team's shortstop. Is there a minor adjustment to the infield model that can account for the 2B/SS proximity, and eliminate the possibility that both the SS and 2B's ability to make plays are affected by the BIPs they can catch? In any case, the results and methods used in this article demonstrate the power of statistical analysis in sport when a statistician is equipped with computational power.

\par The articles covered so far accurately embody the different ways statistical analysis has been performed in sport. In the last paragraph the term computational power was mentioned. If SAFE was proposed 15 years ago, it would have been computationally infeasible. As computational power has gotten cheaper, statistically analysis has benefitted. Another beneficial use of this computational power is to perform statistical analysis using techniques in Artificial Intelligence. The area of Artificial Intelligence contains numerous algorithms, many of which are built on statistical theory, that have the potential to improve statistical analysis in sport. We conclude this review by exploring a case study which applies an Artificial Intelligence technique to football.

\subsection{Modelling using Neuro-Dynamic Programming}

	As we have seen, modelling sports using various statistical models is challenging. A case study performed by Patek and Bertsekas (1996) used Neuro-Dynamic programming (NDP) to simulate the offensive play calling in football~\cite{ndp}. The reason to use football as the ``testbed"  was due to it's state space: it lies on the boundary between medium-scale (tractable) and large-scale (computationally infeasible) problems. The reason this problem isn't computationally infeasible is due to $\mathit{discretization}$ of the yardage, which in the real-world is $\mathit{real-valued}$. 
	\par This discretization is equivalent to the two articles previous because it seeks to represent a real-valued state space accurately, but also efficiently. Similar to the previous articles, sample data that was representative of typical play was generated to ensure that biased approximations wouldn't result. This was performed by using a Gibbs sampler.

\par	A few rules were enforced to create a simplified version of football. Consider a single offensive drive in the middle of a game that's infinitely long. The objective for ``our" team is to maximize the score during ``our" team's offensive drive $\mathit{offset}$ by the rival team's expected score from where they receive the ball. The state of ``our" team is represented by integers $x,y,d$ where $x$ is the number of yards until the goal is reached, $y$ is the number of yards remaining to reach a first down, and $d$ is the down number. The offensive drive for either team will terminate when:
\begin{enumerate*}
\item A team fails to get a new set of downs. That is, they fail to achieve a first down after four plays. For those unfamiliar with American Football: A team always begins its offensive drive with a first down and must achieve $y$ yards in four plays to receive a new first down, and a new $y$ value. If a team doesn't achieve $y$ yards in one play, the down number $d$ is incremented, and $y$ is decremented by the number of yards obtained from this play.
\item Touchdown is scored. (This means the offense must achieve $x \leq 0$.).
\item A turnover either through a fumble on a run attempt, or interception during a pass attempt, or whenever a team elects to punt or attempt a field goal.
\end{enumerate*}
The outcome of a given offensive drive is random, depending on the quarterback's strategy and the associated transition probabilities for the diverse play options and points in the state space. The decision-maker has four play options from which he can choose: 0- Run, 1-Pass (attempt), 2-Punt, 3-Field Goal (attempt). Further information on each type of play option can be obtained by consulting the literature.
\par With these rules and ideas established, the problem of maximizing total expected reward can be represented as the $\mathit{stochastic\;shortest\;path}$ problem because ``maximizing net reward" is equivalent to ``minimizing net costs"; therefore NDP methods mentioned in Appendix~\ref{nddp} are applicable. As mentioned above, there are a finite number of states for which quarterback should have a control action in mind. Such states where the quarterback will have a control action in mind will be denoted by $i \in \mathcal{S}$, where $\mathcal{S}$ is finite.  The triple $(x_i,y_i,d_i)$ denotes the number of yards to the goal, yards to the first down, and the down number corresponding to state $i \in \mathcal{S}$. By abuse of notation, $(x_i,y_i,d_i)$ is written as $(x,y,d)$ because there is only one triple for each state $i \in \mathcal{S}$ and vice-versa. 
\par The quarterback's $\mathit{policy}$ is represented as a function $\mu: \mathcal{S} \to U$, where $U = \{0,1,2,3\}$, the control options available to the quarterback. Whenever possession is lost by ``our" team, we transition to an absorbing state $T$ with zero-reward, similar to transitioning to the three-out state in the Markov Chain model for baseball. The quantity of the reward $g$ is the score received at the end of ``our" team's drive minus the expected score obtained by the opposing team starting at the given field position.
\par With this representation, the optimal policy found it best to run between $x=1$ and $x=65$, attempt to pass from $x=66$ to $x=94$,  and run again from $x=95$ to the goal ($x=100$). The reward function that used this optimal policy showed yielded an expected reward of -0.9449 points when starting from $x=20$. This meant that if the ``our" team received the ball at $x=20$ every time, they would lose the game. This result is a function of arbitrarily-set parameters in the model. If they are adjusted sufficiently, positive reward could be obtained. 
\par What was also interesting was when $y$ was varied for each $x$, because $y$ can be as large as $x$ (although the likelihood of $y > 20$ is very small). On second downs, it was found that the optimal policy dictates pass attempts be made for a wide array of $x,y$ values, with the remaining $x,y$ values as run attempts. For third down, the optimal policy recommended a pass attempts, but if $x$ and $y$ were large enough, the policy suggested punting. For fourth down, the policy had a diverse recommendation; if ``our" team was close enough to the goal or a first down, a running or passing play was recommended. If either a first down or the goal is far away, either a field goal attempt or punt was recommended. By producing a policy that is in agreement with commonly-employed football strategy, Neuro Dynamic Programming shows promise for devising the ``optimal" plan when playing a general opponent. 

\par Particularly advantageous to using NDP techniques was the ability to hypothesize a class of policies that represented legitimate football strategy. These policies can then be simulated. Using a heuristic policy that is reflective of most good ``play-callers" in football, it was found that this policy had an expected reward was -1.26 which is .32 game points $\mathit{worse}$ than optimal when starting from state $i^* \leftrightarrow (x_{i^*} = 80, y_{i^*} = 10, d_{i^*}= 1)$. This is interesting because even though the result is worse than the $\mathit{optimal}$ policy, the fact that an arbitrary heuristic policy could be evaluated suggests the optimal policy could be surpassed with a combination of (unbiased) parameter adjustments and refinement of the policy itself. 

\par We see from these results that, even though this is a simplified version of football, that a Markov representation of sports combined with techniques in Artificial Intelligence (such as NDP) hold promise for analysis in sports. This is shown by finding an optimal policy that is realistic and reflective of play calling in football. In baseball, it is possible that a similar model could be constructed.

\section{Conclusion}\label{con}\
	Throughout this survey, we have been exposed to different methodologies and techniques that all had similar goals: an accurate representation of either players, teams, the performance of players and teams, and performance of players in select situations. 
	\par An alternative methodology for evaluating players is to construct predictive models that are trained on real-world data. NDP produces a promising result, especially considering how it was trained on sampled data that was representative of the real-world. It could be the case that training on real world data produces a better result than the one obtained above. Further experimentation is required in order to determine the validity of applying NDP to sports. 
	\par Aside from the previous article using NDP, the literature for statistical analysis in sports using Statistical Learning techniques is sparse. The computer is not just a computational tool, it possesses the ability to greatly improve statistical analysis in sport when Statistical Learning algorithms, which possess elegant mathematical properties, are used. These elegant mathematical properties are exploited to not only reduce the amount of computation, but provide intricate analyses that further substantiate a statistician's results. Application of these techniques to sports will allow statisticians to investigate complex situations such as the effect of runners on-base on a player's pitching strategy. Investigation of such complex situations in areas of sport may lead to development of precise player metrics that quantify players' decision making abilities.

\appendix

\makeatletter   
 \renewcommand{\@seccntformat}[2]{Appendix~{\csname the#1\endcsname}:\;}
 \makeatother

\section{Absorbing State Markov Chains}\
Assume we are given a simple $n$-state Markov chain $(X_t)_{t=1}^\infty$ and states $\{s_1,\ldots,s_n\}$ with an associated $n \times n$ transition matrix P. The $i$th row and $j$th column of transition matrix P correspond to the probability $P(i,j) = \mathbb{P}(X_t = j | X_{t-1}  = i)$. 
	\par An Absorbing Markov Chain is Markov Chain where there exists at least one $\mathit{absorbing}$ state. An absorbing state is defined as a state which, upon transitioning to, cannot be exited. Mathematically, this is state $s_k$, where $1 \leq k \leq n$, such that $P(k,k) = \mathbb{P}(X_t = s_k | X_{t=1} = s_k)  = 1$. 
	\par If we assume we have more than one absorbing state, then denote the set of these states by $\mathcal{E}$. If we denote the set of non-absorbing states as $\mathcal{S} = \{s_1,\ldots,s_n\} / \{\mathcal{E}\}$, then we can construct matrices $Q$ and $S$ such that matrix $Q$ is of dimension $|\mathcal{S}| \times |\mathcal{S}|$ and matrix $S$ is of dimension $|\mathcal{S}| \times |\mathcal{E}|$.  The matrix S is the transition matrix of  the states in $\mathcal{S}$ to states in $\mathcal{E}$.  Matrix $Q$ is the transition matrix from states in $\mathcal{S}$ to states also in $\mathcal{S}$.
	\par Using these matrices $Q, S$, we can find the probability of reaching a state in $\mathcal{E}$  from states in $\mathcal{S}$ by the following expression:
	\begin{equation} (I - Q)^{-1} S  \end{equation}
	which also yields a $|\mathcal{S}| \times |\mathcal{E}|$ matrix. The entries of this matrix give the probability of entering an absorbing state in $\mathcal{E}$ from a state in $\mathcal{S}$. The matrix $(I-Q)^{-1}$ is referred to as the $\mathit{fundamental\;matrix}$. The $(i,j)$th entry of this matrix is the expected number of periods that the Markov Chain spends in non-absorbing state $j$ given that the chain began in state $i$.
	\par As we will see in section~\ref{prvw}, Absorbing Markov Chains are important for certain representations in the game of baseball.

\section{Gibbs Sampling}
	Some papers in this review will use the idea of Gibbs sampling, which is an algorithm that generates a sequence data samples from a multivariate probability distribution. The point of this generating is to approximate the joint posterior distribution in order to marginalize with respect to some subset of the variables. If we have random variables $X,Y$, then to compute f(x): $$f(x) = \int f(x,y)dy $$ however, in practice this can be a difficult integration. Often in statistical models, either $f(y|x) or f(x|y) $ are available, and the Gibbs sampler generates samples of f(x) from these distributions. The generated sequence of random variables is called a ``Gibbs sequence", and the act of generating this sequence is called ``Gibbs Sampling": 
	
	$$Y_0',X_0',Y_1',X_1',Y_2',X_2',\ldots,Y_k',X_k'$$
	where $Y_0' = y_0'$ is specified and the remaining samples are iteratively generated from:
	\begin{equation}
	\begin{split}
	X_k' ~ f(x |Y_j' = y_j') \\
	Y_{j+1}' ~ f(y | X_j' = x_j')
	\end{split}
	\end{equation}
	\newline
	With the condition that $k$ is large enough, the distribution of $X_k'$ is said to converge to f(x), the ``true" marginal probability distribution function for random variable X, when $k\to  \infty$. Therefore, if we have a large enough sequence of samples, then $X_k'=x_k'$ is said to be a ``true" sample from the distribution of random variable X~\cite{cassella}.

\section{Neuro Dynamic Programming}\label{nddp}
	When dealing with the general class of Markov Decision processes, we are usually interested in finding a policy (probability distribution) such that our long-term total reward is maximized. By applying a control at a given state, the probability distribution that governs the immediate reward and transitions to the next state is determined. Given a starting state $i$, the long term discounted reward is:
	\begin{equation}
		\small J^*(s_0) = \max_{\pi = \{\mu^0, \mu^1,\ldots \} \in \Pi} \mathbb{E}\bigg\{\sum_{k=0}^\infty \alpha^kg(s_k, \mu^k(s_k),s_{k+1}) | s_0, \pi \bigg\} 
		\end{equation}
where:
\begin{itemize*}
\item The states $\{s_k\}$ is a trajectory representing the sequence of states in a $\mathit{finite}$ state space $\mathcal{S}$.
\item $\pi = \{\mu^0, \mu^1,\ldots\} \in \Pi$ is the policy, which is a sequence of functions $\mu^k \in M$ mapping the state space $\mathcal{S}$ to a finite set of ``allowable" controls U.
\item The reward from transitioning from $i$ to $j$ under the control $u$ is $g(i,u,j)$
\item $\alpha \in (0,1]$ is the discounting rate for rewards in the future
\item The expectation over all trajectories of states $\{s_k\}$ that are possible under $\pi$.
\end{itemize*}

When looking at the $\mathit{stochastic\;shortest\;path}$ problem and the $\mathit{discounted\;reward\;problem}$, the optimal reward function from each state is:
\begin{equation}
J^*(i) = \max_{u \in U} \bigg[ \sum_{j \in S} p_{ij}(u)(g(i,u,j) + \alpha J^*(j))\bigg],\;\; \mathit{\forall i\;\in\mathcal{S}}
\end{equation}
This expression is often referred to Bellman's equation. It is the value of the control $u$ which achieves the maximum in Bellman's equation for each state $i \in \mathcal{S}$ and determines the stationary optimal policy $\mu^*$.
\par Because solving explicitly for this system of equations is difficult, $\mathit{value\;iteration}$ is used. The goal of value iteration is to start with a guess for $J^*$, called $J^0$, defined for all states $i \in \mathcal{S}$. With successive iterations, the function for the $k$th iteration is:
\begin{equation}
J^k(i) = \max_{u \in U} \bigg[ \sum_{j \in S} p_{ij}(u)(g(i,u,j) + \alpha J^{k-1}(j))\bigg],\;\; \mathit{\forall i\;\in\mathcal{S}}
\end{equation}
where, in the limit $J^* = \lim_{k\to\infty}J^k(i)\;\; \forall i \in \mathcal{S}$. Intuitively speaking, this means the approximation of each state's reward function should approach the value of the optimal reward function $J^*$ as the number of iterations increase.
\par An alternative way to compute the optimal reward function, $J^*$ uses the $\mathit{policy\;iteration}$ algorithm. This algorithm starts with a policy $\mu^0$ and evaluates $J^{\mu^0}$ for all states $i \in \mathcal{S}$. The $k$th iteration for this policy is computed by:
\begin{equation}
\mu^k(i) = \mathrm{arg}\max_{u \in U} \bigg[\sum_{j \in \mathcal{S}} p_{ij}(u) (g(i,u,j) + \alpha J^{\mu^{k-1}}(j))\bigg]\;\;\mathit{\forall i \in \mathcal{S}}
\end{equation}
Which will converge to $J^*$ as long as the evaluations of $J^{\mu^k}$ are exact and $\mathcal{S}$ and U are finite. It is assumed that when these algorithms are used in this article that the discount factor $\alpha = 1$.

	\section{Baseball Terminology}
	We provide a quick description of many terms that are frequently used in baseball.
	\begin{itemize*}
	\item An at-bat is a plate appearance for a batter. This at-bat has a count starting at 0-0. The first number represents the number of ``balls" (explained below), of which there can be a maximum of four. If four ``balls" are achieved, the batter is advanced to first base (called a ``base on ball"). The second number represents the number of strikes, if three strikes are obtained then the batter is out and the plate appearance ends.
	\item A ``ball" is when the pitcher fails to throw a pitch in the batter's strike zone.
	\item Base on balls will be referred to as ``walks" for this paper.
	\item A ``flyout" is when the batter successfully hits the pitch but it was caught by an outfielder on the opposing team.
	\item A ``groundball" is when the batter hits the pitch but it rolls on the ground.
	\item A ``sacrifice" is when the batter hits a pitch that results in either a flyout or groundout to advance a runner on base (which includes scoring a runner on third base).
	\item An inning in baseball consists of each team taking their turn to bat. Each inning for a team consists of three outs, and therefore at least three batters will get a plate appearance prior to the other team taking their turn to bat.
	\end{itemize*}
	\onecolumn

\end{document}